\documentclass[a4paper]{jpconf}

\usepackage{graphicx}
\usepackage{subfigure}
\usepackage{mathrsfs}
\usepackage{amsmath}
\usepackage{slashed}
\usepackage{setspace}
\usepackage{footnote}
\usepackage{comment}
\usepackage[hyperindex,breaklinks]{hyperref} 

\usepackage{bm}        
\usepackage{amssymb}   
\usepackage{rotating}
\usepackage{xspace}

\begin{document}

\title{TOP 2014: Experimental Summary}

\author{Christian Schwanenberger}

\address{The University of Manchester, Oxford Road, Manchester, M13
  9PL, UK; \\
 now: Deutsches Elektronen-Synchrotron (DESY), Notkestr. 85,
  D-22607 Hamburg, Germany}

\ead{christian.schwanenberger@cern.ch}

\newcommand{\alphalp}{\ensuremath{\alpha_\ell P}}
\newcommand{\alphaip}{\ensuremath{\alpha_i P}}
\newcommand{\alphalpcpc}{\ensuremath{\alpha_\ell P_{\mathrm{CPC}}}}
\newcommand{\alphalpcpv}{\ensuremath{\alpha_\ell P_{\mathrm{CPV}}}}
\newcommand{\ljets}{$\ell+$jets}
\newcommand{\ejets}{$e+$jets}
\newcommand{\zjets}{$Z/\gamma^*$+jets}
\newcommand{\mujets}{$\mu+$jets}
\newcommand{\mcatnlo}   {\mbox{\textsc{MC@NLO}}}
\newcommand{\pythia}   {\mbox{\textsc{Pythia}}}
\newcommand{\herwig}   {\mbox{\textsc{Herwig}}}
\newcommand{\jimmy}   {\mbox{\textsc{Jimmy}}}
\newcommand{\alpgen}   {\mbox{\textsc{Alpgen}}}
\newcommand{\powheg}   {\mbox{\textsc{Powheg-Box}}}
\newcommand{\sherpa}   {\mbox{\textsc{Sherpa}}}
\newcommand{\madgraph}  {\mbox{\textsc{Madgraph}}}
\newcommand{\acer}   {\mbox{\textsc{Acer}}}
\newcommand{\ppm}{$\pm$}
\newcommand\mll{\ensuremath{m_{\ell\ell}}}
\newcommand{\etmiss}{$\et^{\textrm{miss}}$}
\newcommand\htlep{\ensuremath{\HT}}
\newcommand\dbq{$\Delta\phi(\ell,b)$}
\newcommand\ddq{$\Delta\phi(\ell,d)$}

\newcommand{\vect}[1]{\hat{#1}}
\newcommand{\dphi}{$\Delta \phi$}

\def\ee{\ensuremath{e^+ e^-}}
\def\mumu{\ensuremath{\mathrm{\mu^+ \mu^-}}}
\def\emu{$e^{\pm} \mu^{\mp}$}

\newcommand\T{\rule{0pt}{+3ex}}
\newcommand\B{\rule[-2ex]{0pt}{0pt}}

\newcommand\TT{\rule{0pt}{+2.3ex}}
\newcommand\BB{\rule[-1.5ex]{0pt}{0pt}}

\newcommand\TTT{\rule{0pt}{3ex}}
\newcommand\BBB{\rule[-2ex]{0pt}{0pt}}

\newcommand{\ahel}{$A_{\rm helicity} = 0.38 \pm 0.04$}
\newcommand{\fsm}{1.20 $\pm$ 0.05 (stat) $\pm$ 0.13(syst)}

\newcommand{\mtop}{\mbox{$m_t$}}
\newcommand{\stopi}{\mbox{$\tilde{t}_1$}}
\newcommand{\stopibar}{\mbox{$\bar{\tilde{t}}_1$}}
\newcommand{\stopistopibar}{\mbox{$\tilde{t}_1\bar{\tilde{t}}_1$}}

\newcommand{\xiz}{\mbox{$\tilde{\chi}_1^0$}}
\newcommand{\mstop}{\mbox{$m_{\tilde{t}_1}$}}
\newcommand{\xstop}{\mbox{$\sigma_{\tilde{t}_1\tilde{t}_1}$}}
\newcommand{\rstop}{\mbox{$R_{\tilde{t}_1\tilde{t}_1}$}}
\newcommand{\mxiz}{\mbox{$m_{\tilde{\chi}_1^0}$}}

\newcommand{\mstopobs}{191} 
\newcommand{\mstopexp}{178} 

\newcommand{\Wt}{\ensuremath{\mathit{\!Wt}}\xspace} 

\newcommand{\ttbar}     {\mbox{$t\bar{t}$}\xspace}
\newcommand{\ppbar}     {\mbox{$p\bar{p}$}\xspace}

\begin{abstract}
A summary of the experimental results of the TOP2014 International
Workshop in Cannes, France, is 
presented. This inspiring conference clearly showed the richness and
diversity of top-quark physics research. Results cover a very broad
spectrum of 
analyses involving studies of the strong and
electroweak interactions of the top quark, high-precision measurements
of intrinsic top-quark properties, developments of new tools in
top-quark analyses, observations of new Standard Model processes,
the interaction between the top quark and the Higgs boson and
sensitive searches for new physics beyond the Standard Model.
\end{abstract}

\section{Introduction}
The top quark, discovered in 1995 at the Fermilab $p\bar{p}$
collider Tevatron, is the most recent member of the known families of
quarks. The Large Hadron $pp$ collider (LHC) is a top-quark factory
providing high-precision access 
to top-quark physics. Although this particle appears to have no
substructure and to be pointlike, its mass is of the order of a gold
nucleus. Therefore it is believed to be closely related to the mechanism
of mass generation together with the Higgs boson. Thus, to test our
current theory of elementary particles in detail, it is of particular
importance to measure the properties of the top quark precisely, to
analyze its strong and electroweak interactions and to investigate its
interaction with the Higgs boson. Because of its potentially special role,
the top quark might be exotic in some way, and could offer a first
glimpse of physics beyond the Standard Model (SM). Therefore, it is important to
perform both direct searches in the top-quark sector and indirect
searches for new physics through high-precision measurements of
top-quark cross sections and properties. In this
proceeding recent results from the Tevatron collider at a
center-of-mass energy (CME) of 1.96~TeV and from the LHC at a CME of 7
and 8~TeV are highlighted. 

\section{Top-quark-pair production and QCD studies}
\begin{figure*}[htpb!]
\vspace*{-2.3cm}
\begin{center}
\subfigure[]{\includegraphics[width=0.51\textwidth]{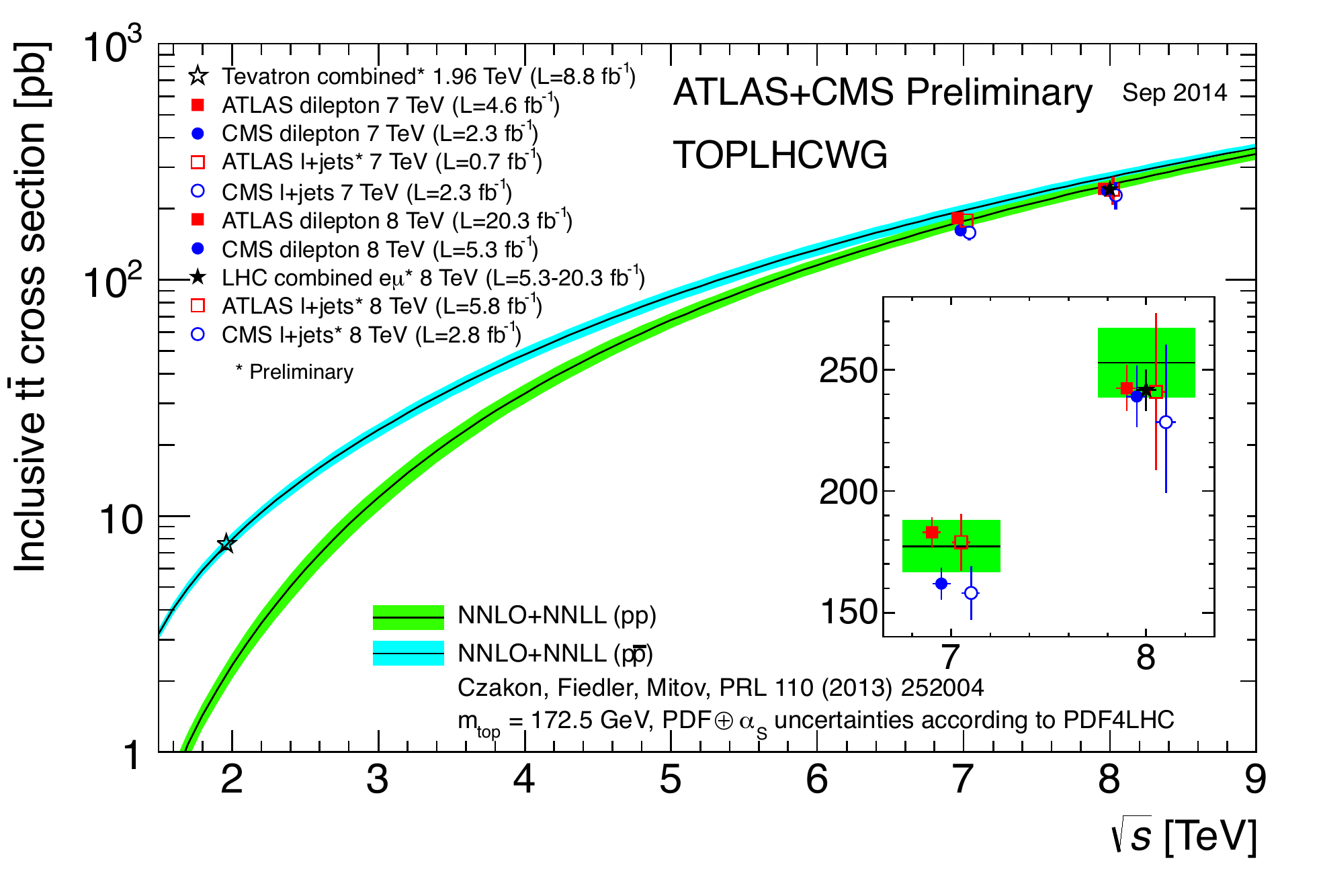}}\label{fig:tt_xsec_colliders}\hspace{0cm}
\subfigure[]{\includegraphics[width=0.45\textwidth]{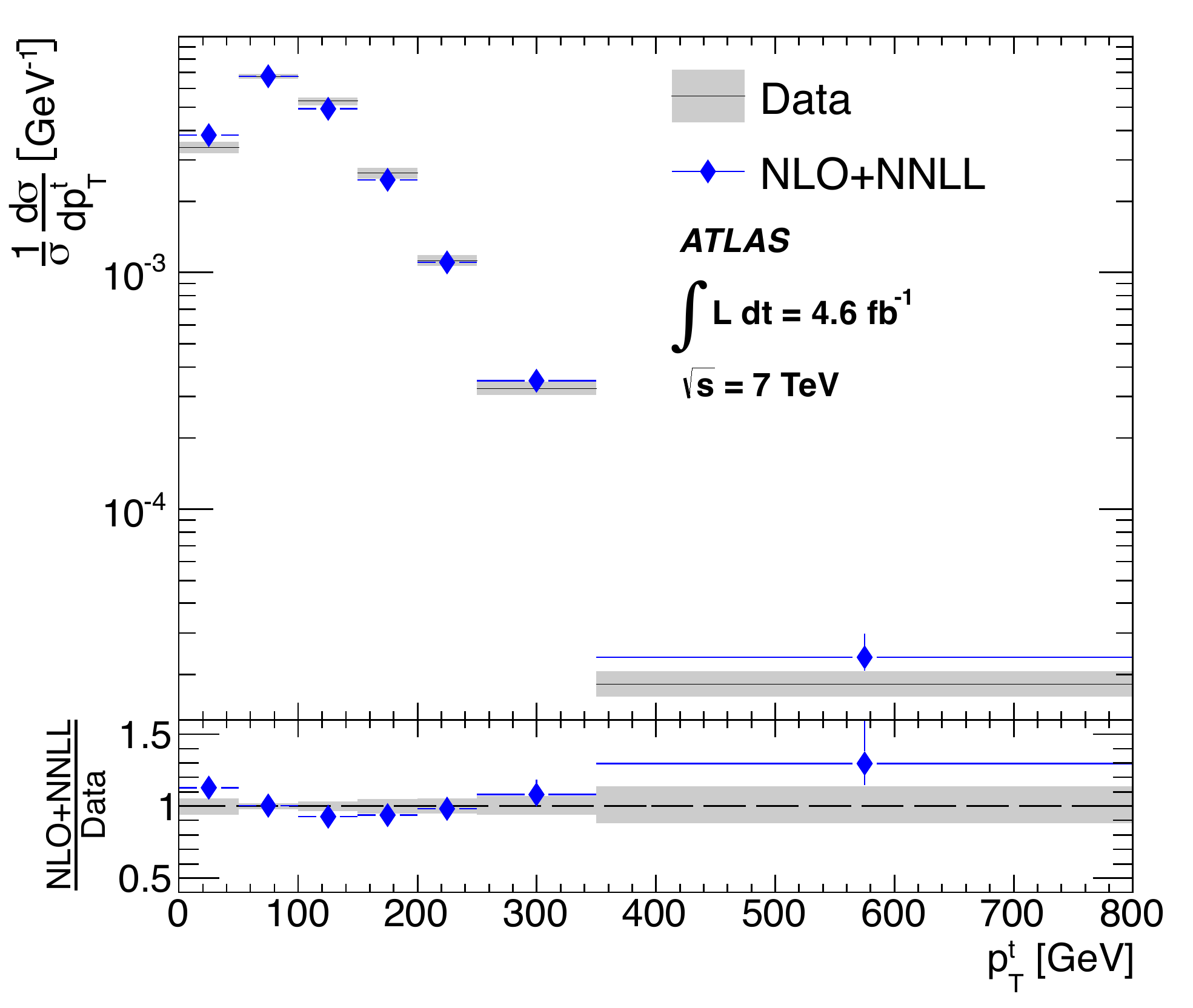}}\label{fig:top_pt_ATLAS}\\
\subfigure[]{\includegraphics[width=0.45\textwidth]{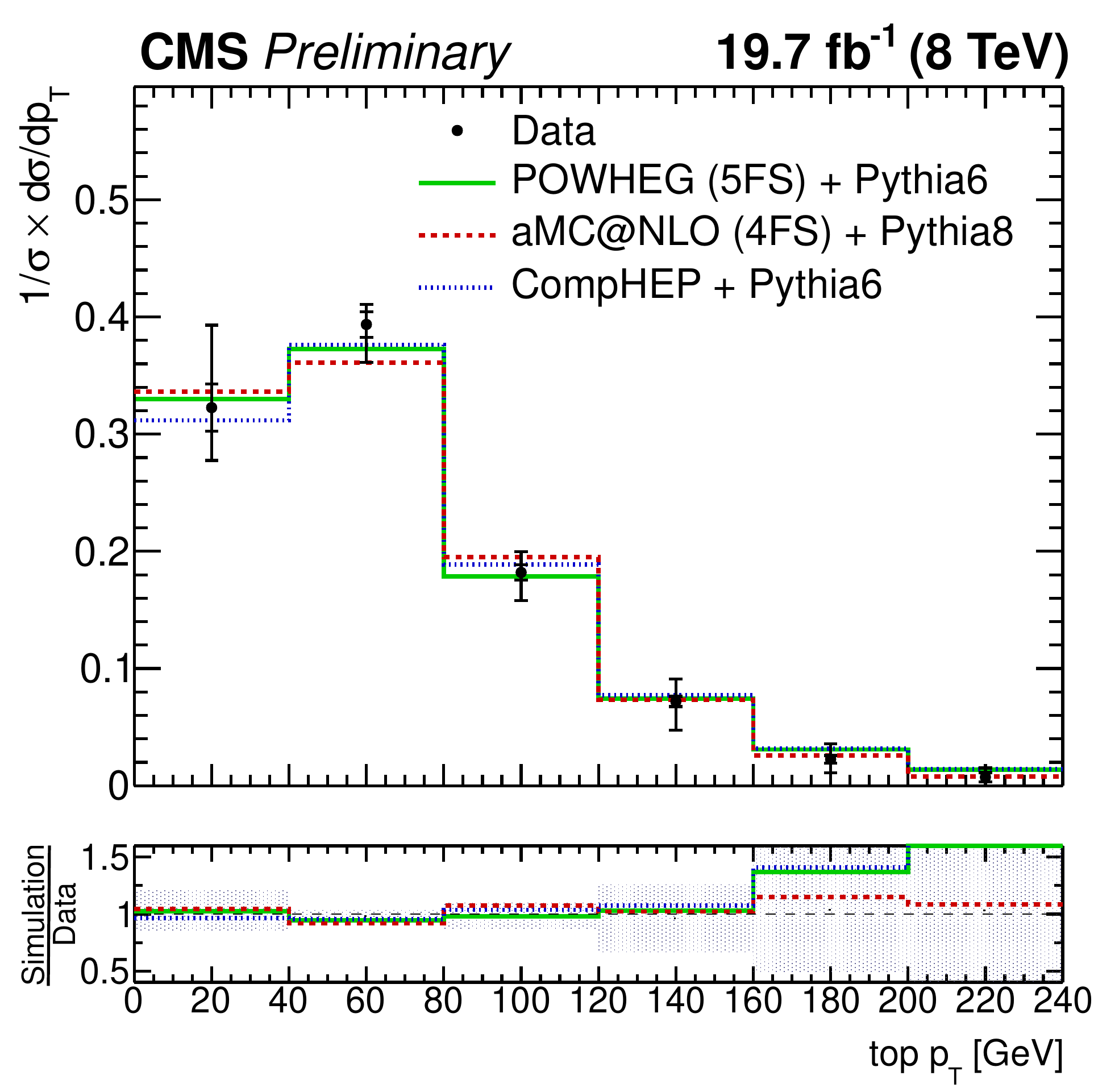}}\label{fig:top_pt_CMS}\hspace{1.5cm}
\subfigure[]{\includegraphics[width=0.42\textwidth]{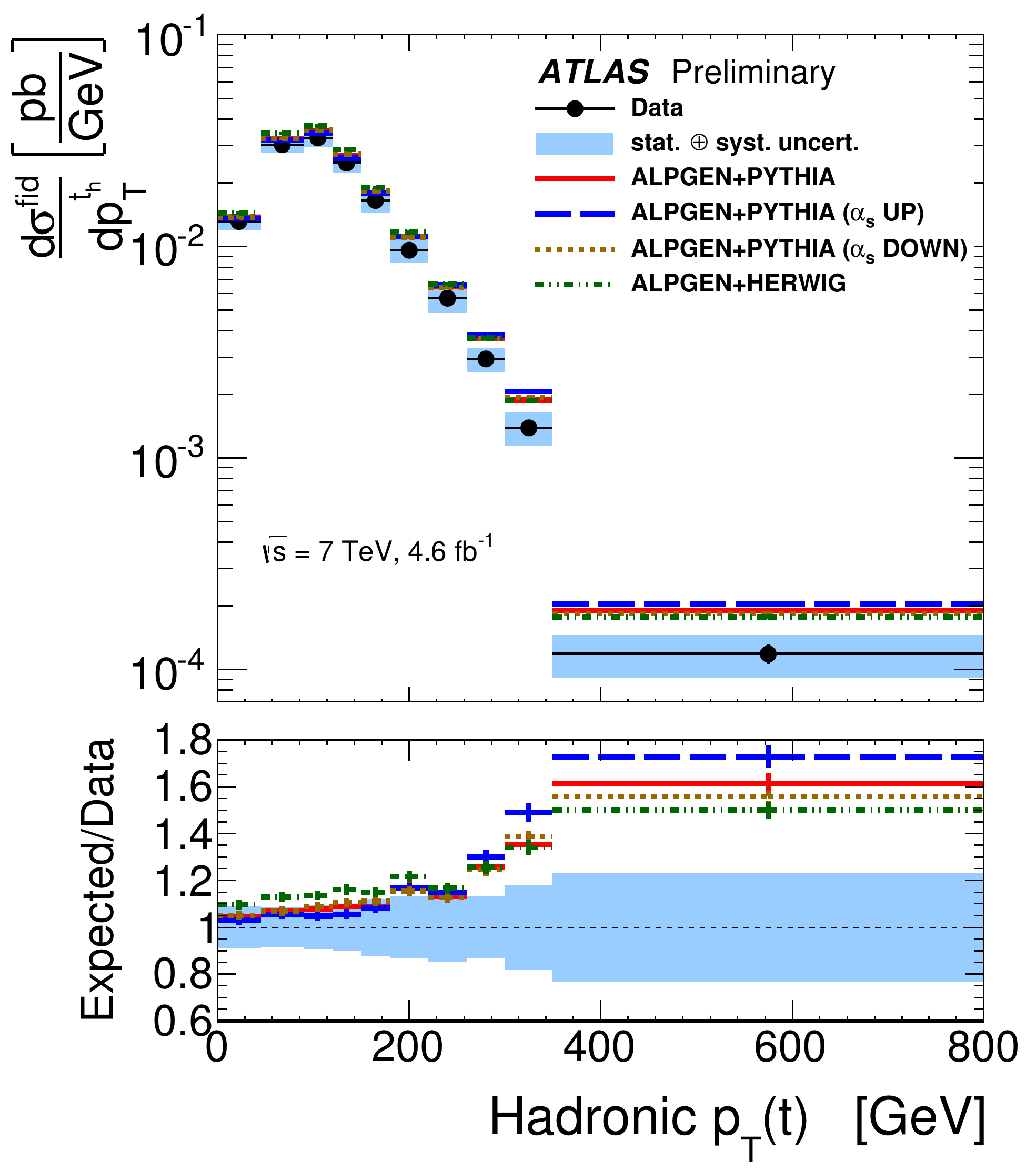}}\label{fig:top_pt_pseudo_top}\\
\subfigure[]{\includegraphics[width=0.38\textwidth]{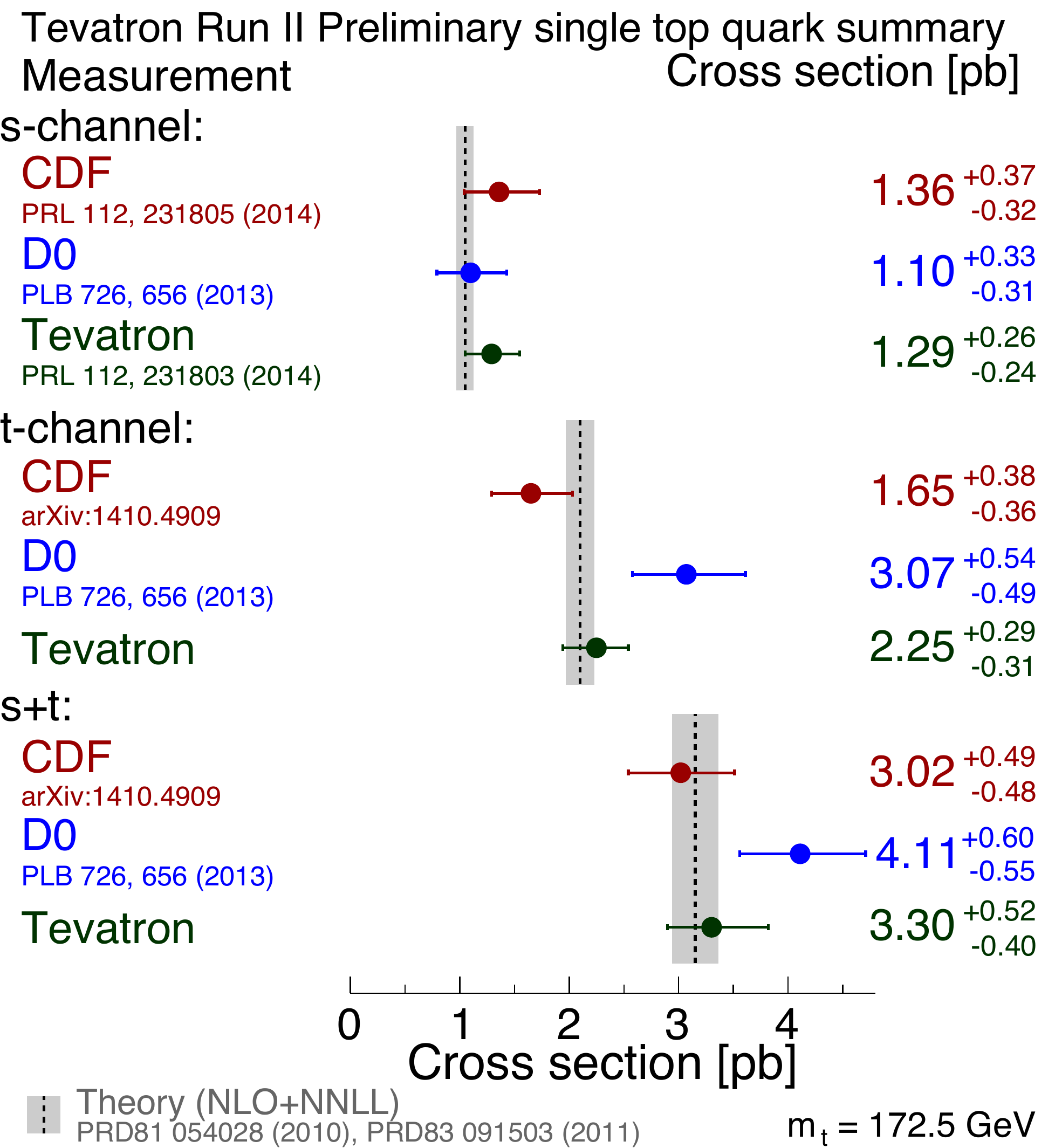}}\label{fig:sitop_Tev}\hspace{1.1cm}
\subfigure[]{\includegraphics[width=0.50\textwidth]{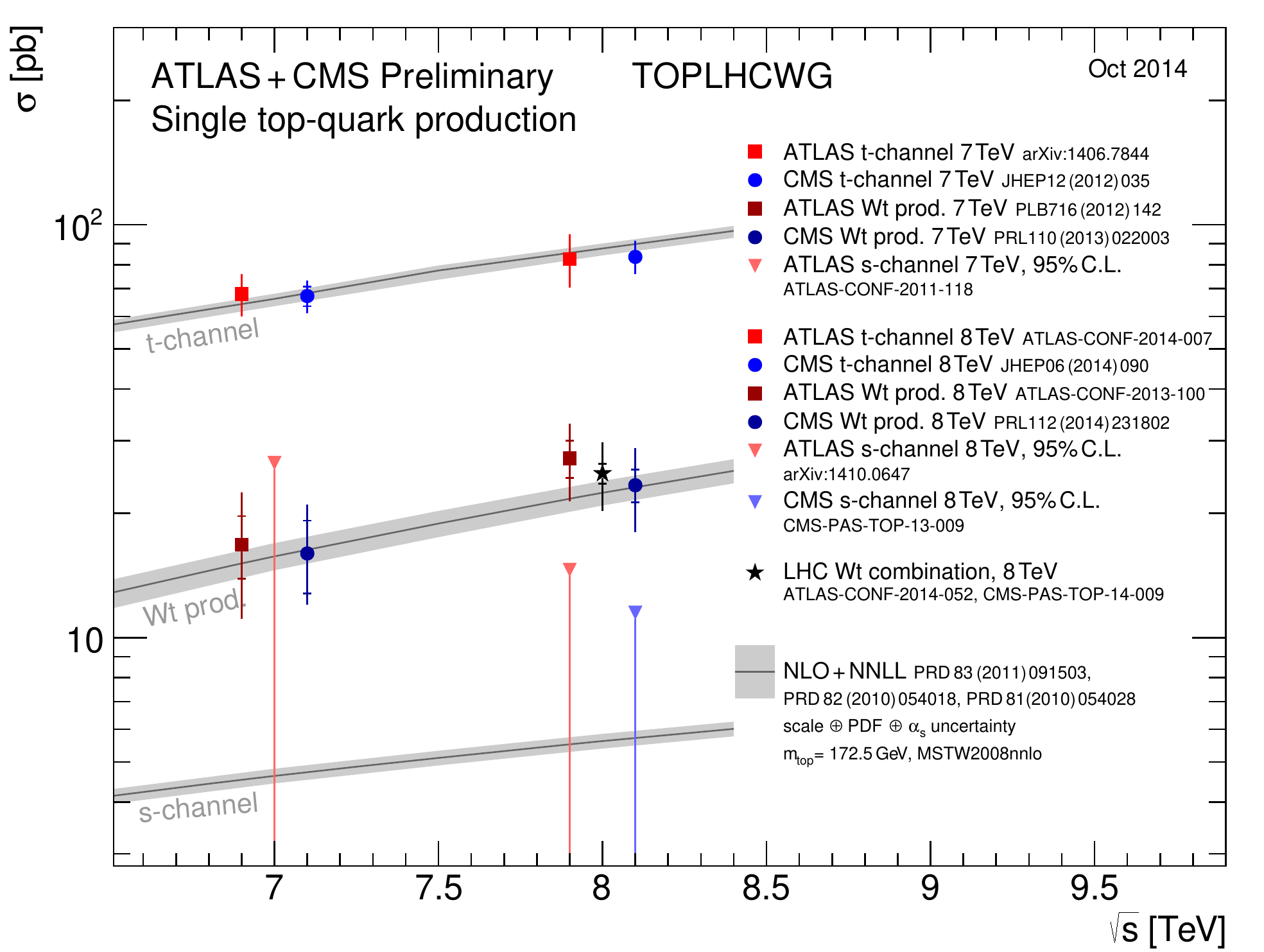}}\label{fig:sitop_LHC}\\
\end{center}
\caption{Top-quark production: (a) Summary of LHC and Tevatron
  measurements of the total \ttbar\ cross section as a function of the
  CME compared to the NNLO+NNLL QCD calculation; (b)
  normalized differential cross sections for the transverse momentum
  of the hadronically decaying top-quark $p^t_T$ at 7 TeV (\ttbar) and (c) at 8
  TeV (single top quark) and (d) for hadronic pseudo-top quarks
  at 7 TeV (\ttbar); (e) single-top-quark production
  cross sections in various channels at the
  Tevatron at 1.96~TeV and (f) as a function of the CME at the LHC compared to NLO+NNLL QCD
  calculations.}
\label{fig:prod}
\end{figure*}

At the Tevatron and the LHC, top quarks are
produced in pairs via the strong interaction and singly via the
electroweak interaction. It is interesting to measure the total
production cross sections in all possible final states, since the impact of potential new
physics could be different in different channels. In top-quark-pair
production all possible final states involving leptons, jets and 
missing transverse momentum have been measured by the two colliders except for final states
with two hadronically decaying tau leptons~\cite{ttbar_xsec}.
The results are in agreement with the predictions in
next-to-next-to-leading order (NNLO) quantum chromodynamics (QCD) including
next-to-next-to leading log (NNLL) corrections~\cite{xsec_tt}. The different
measurements are shown as a function of the CME in
Fig.~\ref{fig:prod}a.

More elaborate tests of QCD than these extrapolated inclusive cross-section
measurements are performed by measuring differential distributions~\cite{xsec_diff}. To
make these measurements more usable for comparisons with calculations
both today and in the future, it is advisable to unfold the
differential distributions and to give the results also explicitly for the
selected phase space where the measurement has actually been
performed.  
The distributions in data provide important information when comparing them to
QCD calculations and Monte Carlo (MC)
simulations. As an example, Figs.~\ref{fig:prod}b and~\ref{fig:prod}c
show the distribution of the top-quark transverse momentum
in \ttbar\ production and single top quark $t$-channel
production, respectively. While the data is in agreement with the
predictions of MC simulations for single-top-quark production, for
\ttbar\ production in measurements of both LHC collaborations 
the data distribution is softer than predicted for higher values of the
top-quark transverse momentum~\cite{xsec_diff}.

It is particularly interesting to study differential cross sections in terms of kinematic
variables of a top-quark proxy referred to as the ``pseudo-top quark''
whose dependence on theoretical models is minimal. Observables of pseudo-top quarks are constructed from
objects that are directly related to detector-level
observables. Figure~\ref{fig:prod}d shows the differential
cross section as a function of the hadronically decaying
pseudo-top-quark transverse momentum. Differences between data and 
simulations similar to those in Fig.~\ref{fig:prod}b are observed and still have
to be understood. 

\section{Single-top-quark production and electroweak couplings}
Single top quarks are produced in $p\bar{p}$ and $pp$ scattering at
tree level via the electroweak interaction in the 
$t$-channel via the exchange of a space-like virtual $W$ boson between
a light quark and a bottom quark, in the $s$-channel via the decay of
a time-like virtual $W$ boson produced by quark-antiquark
annihilation, which produces a top quark and a bottom quark, or in
association with a $W$ boson (\Wt). Recently the Tevatron
collaborations have observed $s$-channel
production~\cite{sitop_tev}, ATLAS has found evidence for $\Wt$ associated
production, followed by an observation of $\Wt$ production at the
CMS experiment~\cite{sitop_lhc}. Now all different single-top-quark production subchannels are
observed. Figure~\ref{fig:prod}e and~\ref{fig:prod}f  show the cross-section
measurements of the Tevatron and the LHC, respectively, for different
subchannels which are all in agreement with 
the SM predictions~\cite{sitop-kidonakis}. 

Other SM processes which have not yet been observed but which are
important since they are sensitive to the electroweak $tZ$ and $tW$
couplings are $t\bar{t}Z$ and $t\bar{t}W$ production. The most sensitive search
channels involve three leptons  
($t\bar{t}Z$) and two same-sign leptons ($t\bar{t}W$) in the final
state. Evidence for both processes has recently been claimed at the
LHC using the full 8~TeV datasets~\cite{ttZ_ttW}. The extracted cross
sections are in agreement with the SM predictions. 

\section{Systematic uncertainties and new tools in top-quark physics}
There are many new developments to understand systematic uncertainties in top-quark
measurements presented at this conference~\cite{systs}. For
example, new investigations for a better understanding of the
energy scale of jets induced by $b$ quarks, of the uncertainties of
parton density functions and of modeling uncertainties in MC
simulations have been discussed in detail. The large efforts to
combine many different measurements between the LHC
experiments~\cite{combi_lhc} and 
between the LHC experiments and the Tevatron
experiments~\cite{combi_world}, has led to a very 
fruitful collaboration among all experiments which is indispensable
for a detailed understanding of systematic uncertainties in
top-quark analyses in general.

There is also huge progress in the developments of new tools in
top-quark physics~\cite{tools}. As an 
example, top-quark taggers in boosted top-quark regimes have made fast
progress moving quickly from the R\&D phase to highly sophisticated
performance studies to optimize top-quark-tagging efficiency and
purity. They are utilized, for example, in searches for \ttbar\
resonances (see Fig.~\ref{fig:searches}c) and differential
cross-section measurements for 
highly-boosted top quarks~\cite{boosted}. 

\section{Top-quark properties}
High precision measurements of top-quark properties are
essential since they could be the key to find new physics in the top-quark
sector by comparing the results to the SM predictions.

The top-quark mass is a free parameter in the SM. High accuracy in the
measurement of the top-quark mass is very important since in
combination with the measurements of the
mass of the $W$ boson and the mass of the Higgs boson the self-consistency of
the SM can be tested. The current world average combining measurements
from the Tevatron and the LHC gives $m_t = 173.34 \pm
0.76$~GeV ($\pm 0.44\%$)~\cite{combi_world}. Since then, new measurements have been
performed that are not yet included in the world average. The most
precise new measurements are from the D0
Collaboration, $m_t = 174.94 \pm 0.76$~GeV ($\pm 0.44\%$)~\cite{mass_d0}, and the CMS
Collaboration, $m_t = 172.04 \pm 0.77$~GeV ($\pm
0.45\%$)~\cite{mass_cms}. Investigations to understand the
tension between the these two results are ongoing. Detailed theoretical investigations
of how the measured masses calibrated by different MC simulations translate
into quantum-field-theoretically well-defined mass parameters are
ongoing, as well as studies of how $WbWb$
off-shell effects can affect the measurements. 

\begin{figure*}[htpb!]
\vspace*{-2cm}
\begin{center}
\subfigure[]{\includegraphics[width=0.45\textwidth]{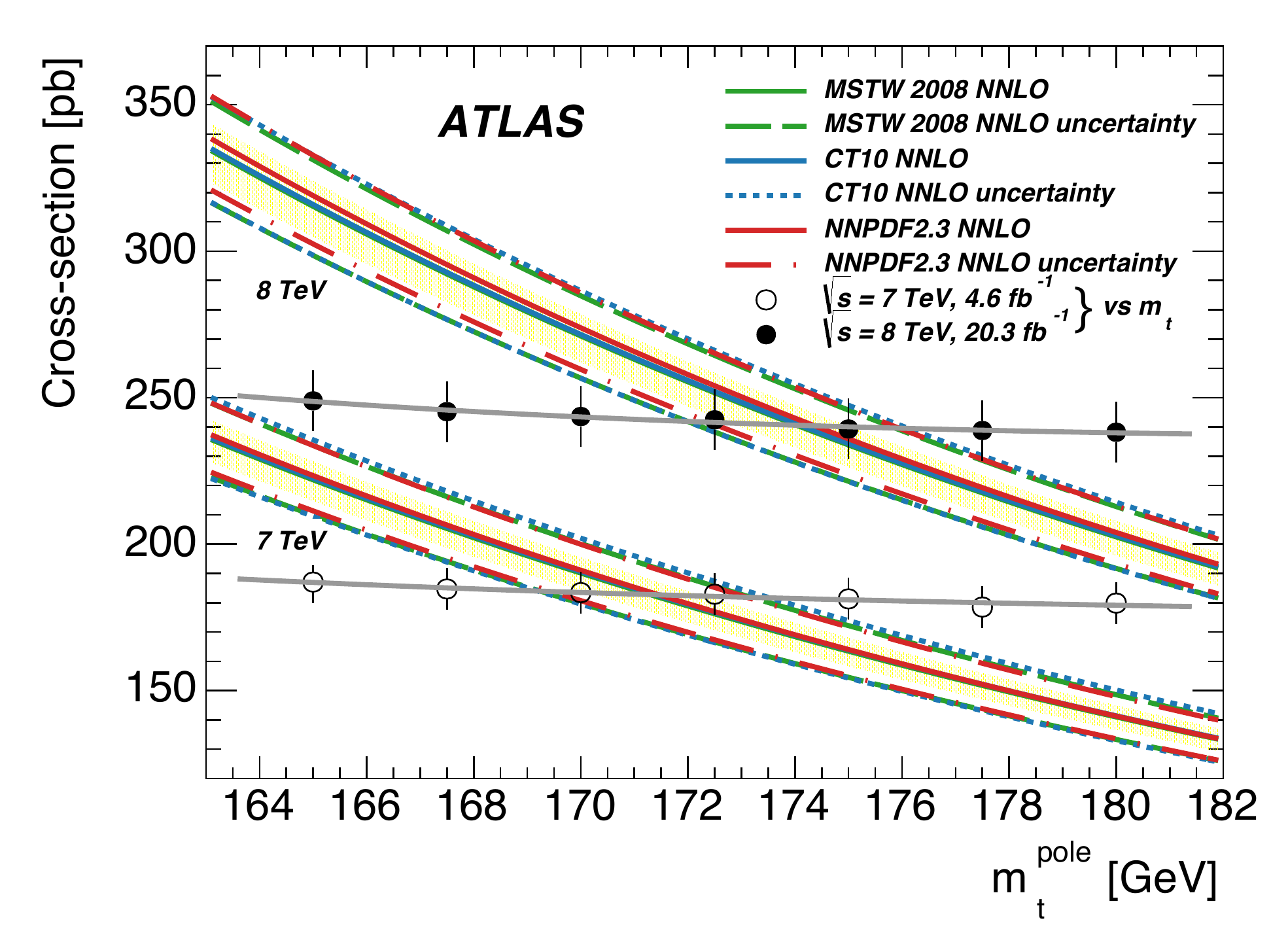}}\label{fig:ttbar_pole}\hspace{0.5cm}
\subfigure[]{\includegraphics[width=0.43\textwidth]{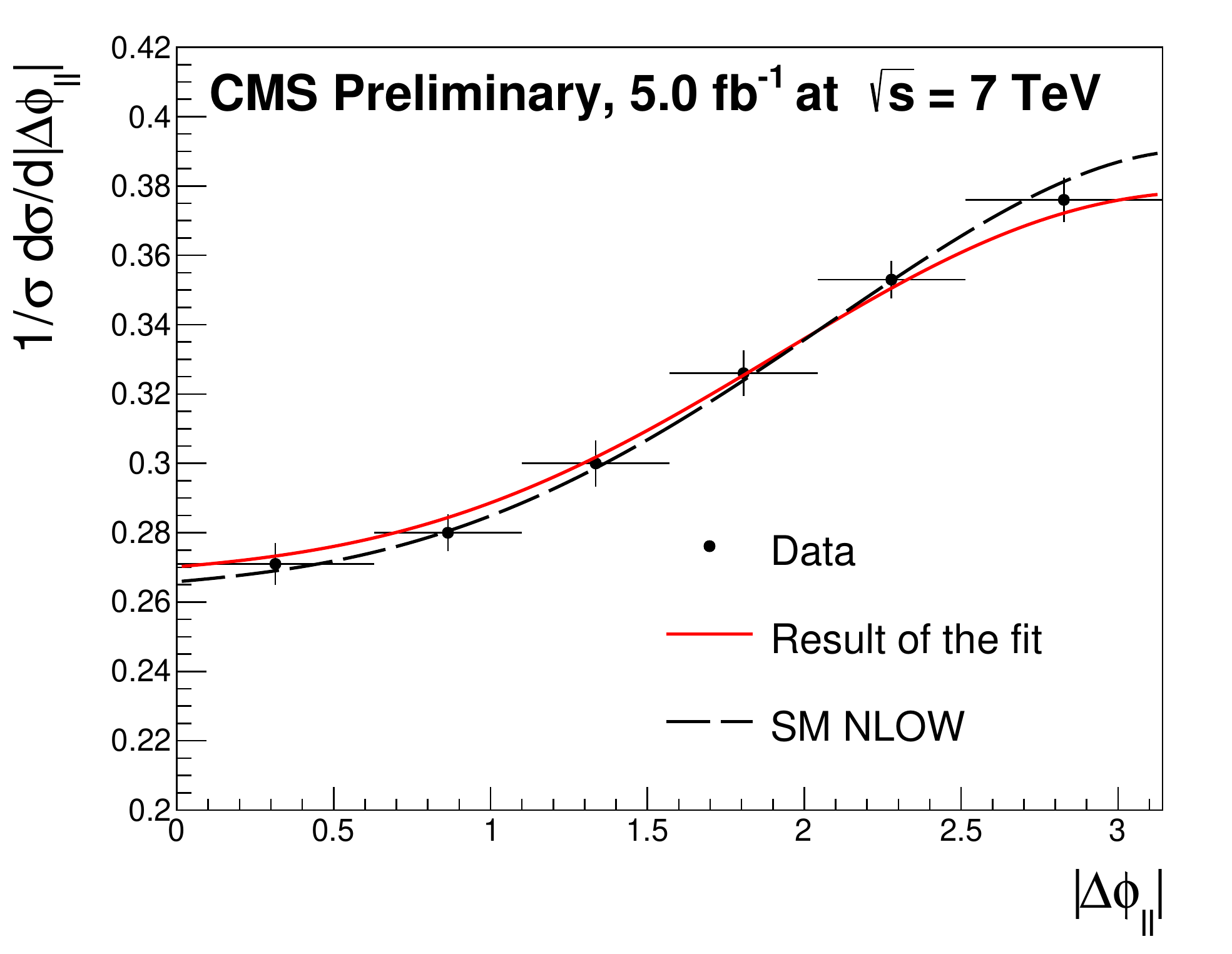}}\label{fig:ttspin_CMS}\\
\subfigure[]{\includegraphics[width=0.35\textwidth]{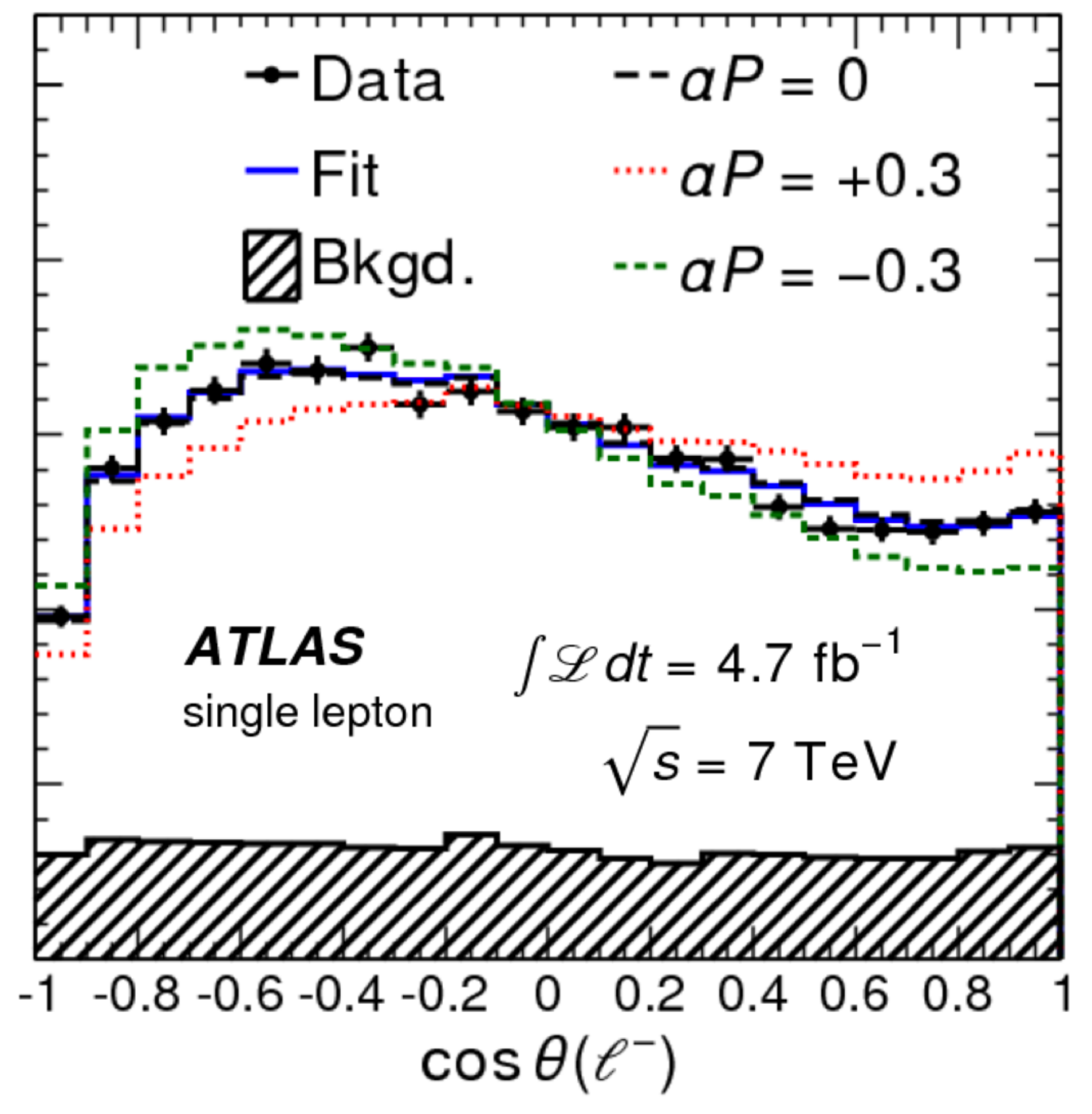}}\label{fig:pol_ATLAS}\hspace{1.5cm}
\subfigure[]{\includegraphics[width=0.39\textwidth]{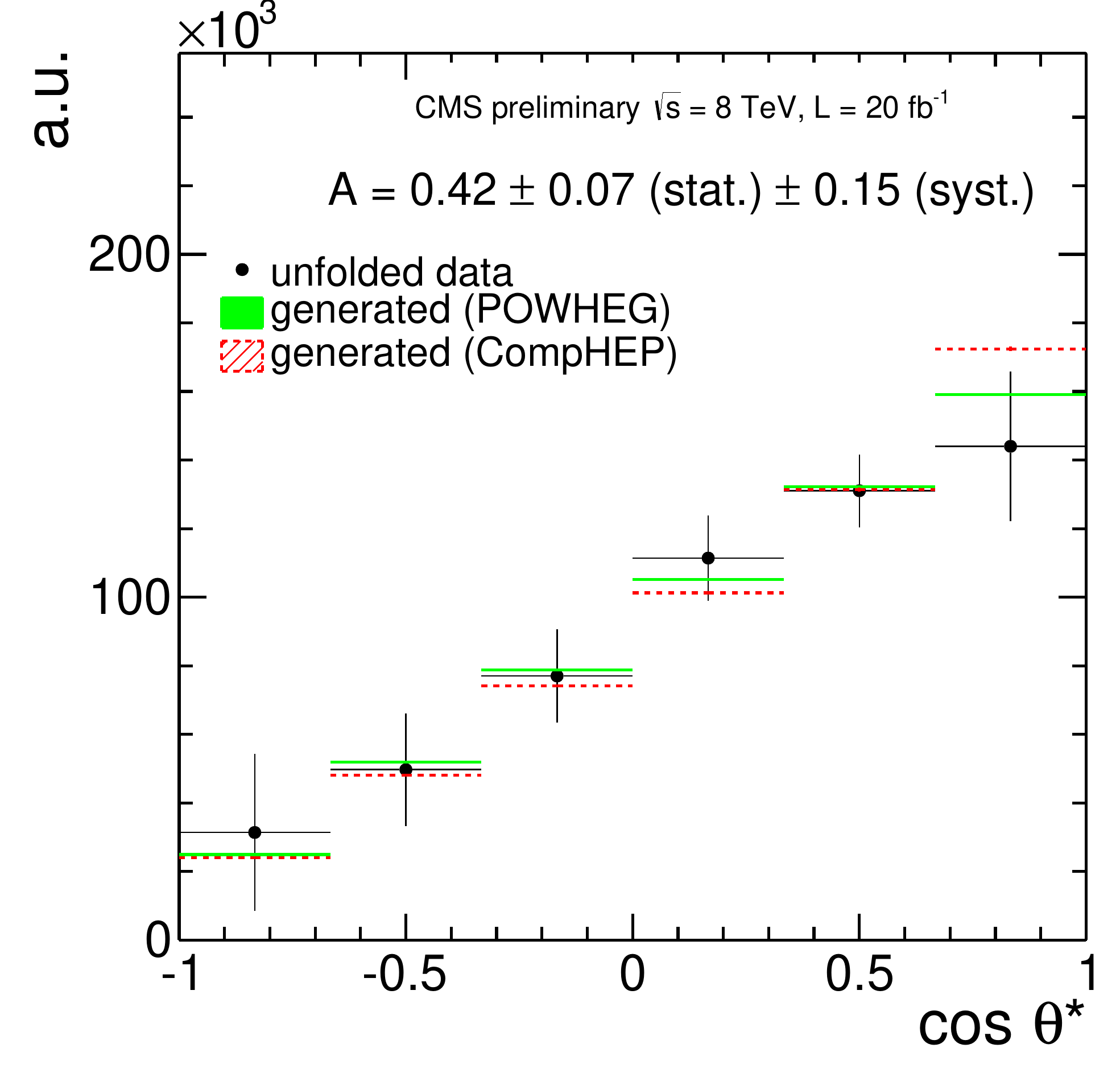}}\label{fig:pol_CMS}\\
\subfigure[]{\includegraphics[width=0.39\textwidth]{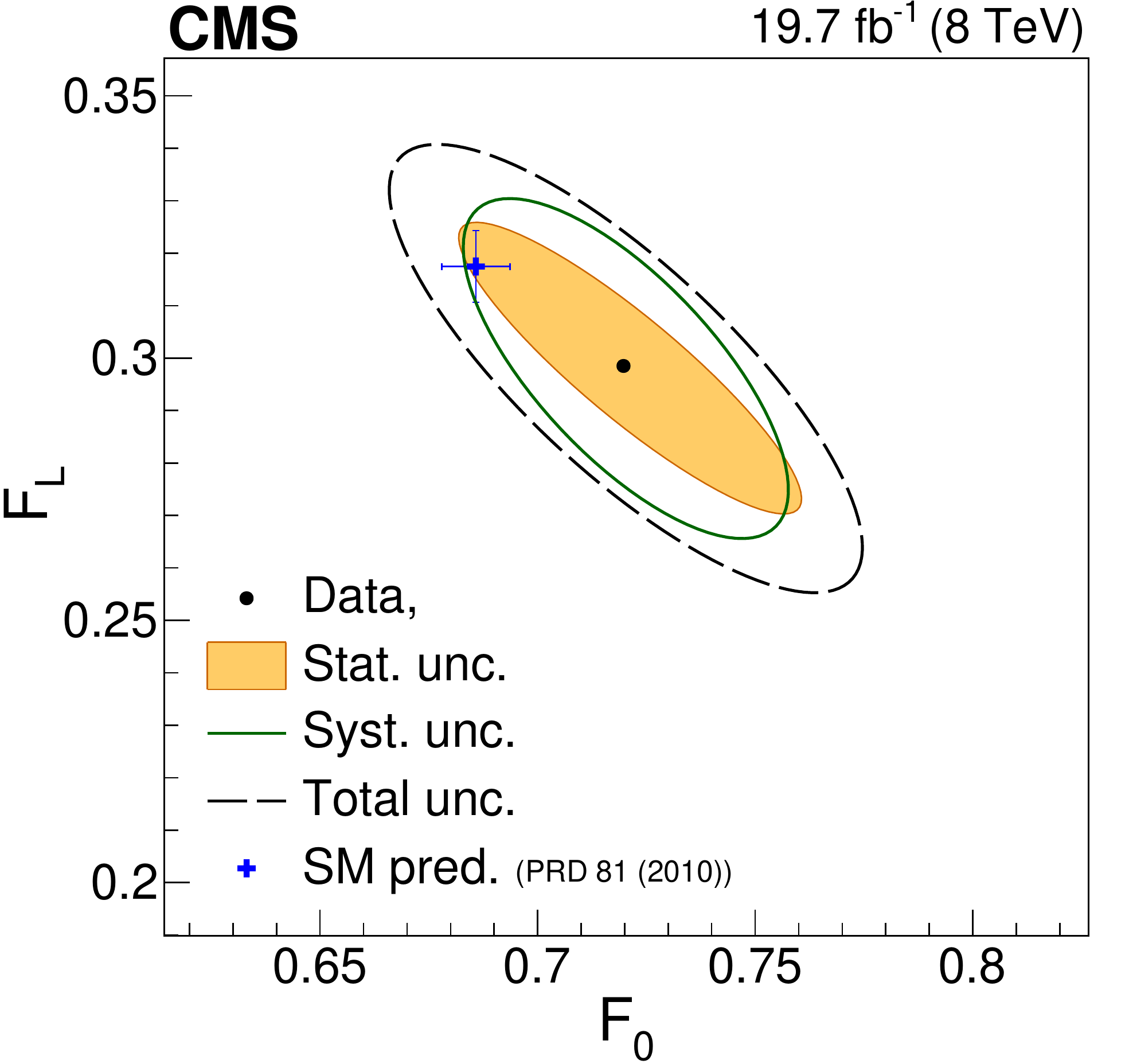}}\label{fig:Whel_CMS}\hspace{1.7cm}
\subfigure[]{\includegraphics[width=0.39\textwidth]{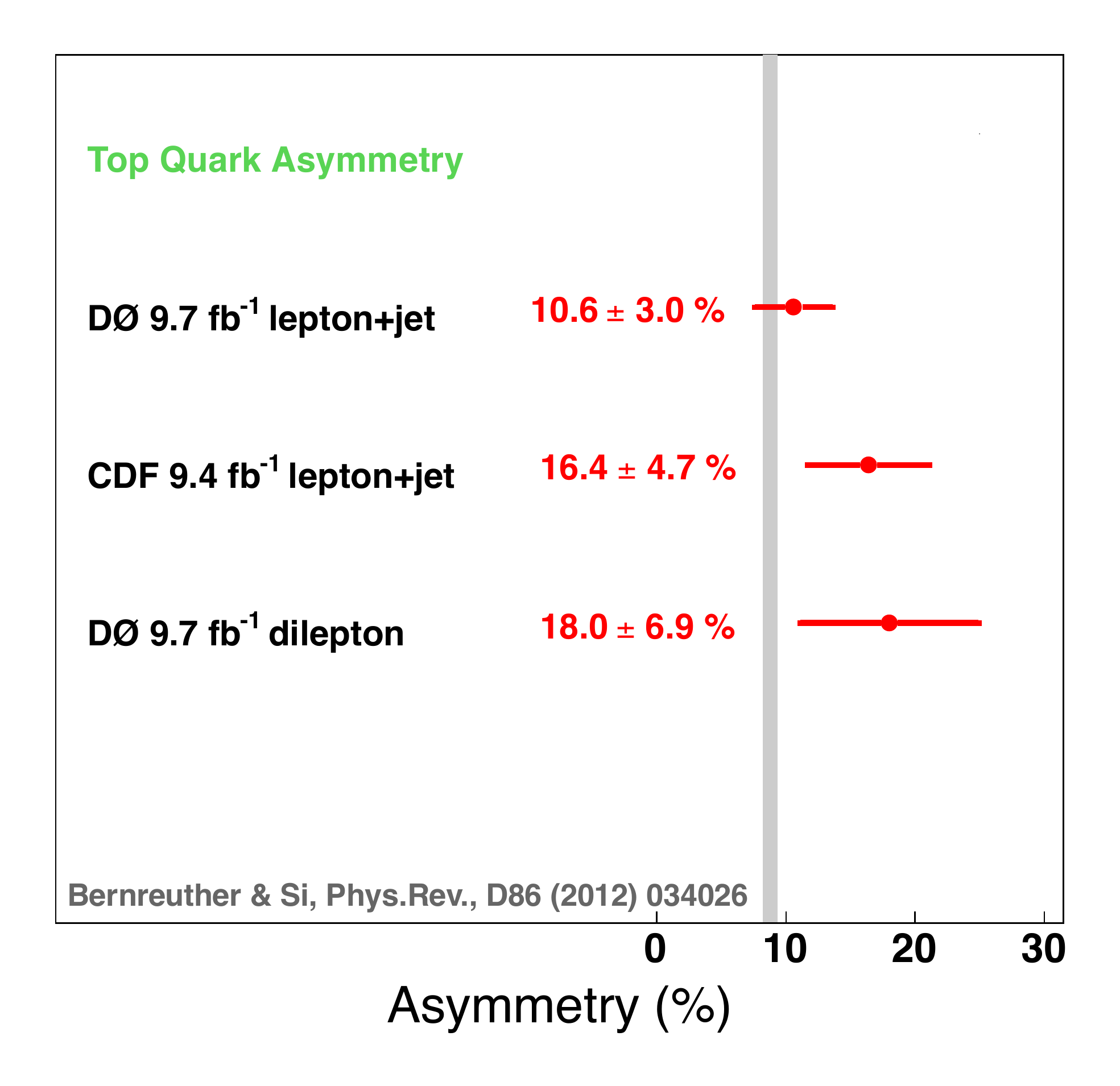}}\label{fig:ttasym_sum}
\end{center}
\caption{Top-quark property measurements: (a) $\sigma_{t\bar{t}}$ at 7 and 8
  TeV as a function of the assumed value of $m_t$ compared to NNLO+NNLL
  predictions as a function of $m_t^{\rm pole}$; (b)
the differential cross section as a function of the difference in
azimuthal angle of the two leptons in \ttbar\ dilepton production as
observed in data and predicted by the SM (dashed curve) at 7~TeV;
(c) top-quark polarization distribution compared to the
  SM prediction of zero polarization and to non-vanishing polarization
  hypotheses in \ttbar\ $e^-,\mu^-$+jets final states at 7~TeV; (d) unfolded
  distribution of the top-quark polarization in single $t$-channel top-quark production
  at 8~TeV in data and SM predictions; (e) extracted left-handed and longitudinal $W$ boson helicity
  fractions from single-top-quark events at 8 TeV, shown as 68\% contours for statistical, systematic, and
  total uncertainties, compared to the SM prediction; (f) summary of \ttbar\ forward-backward asymmetry
  measurements using the full Tevatron dataset.}
\label{fig:properties}
\end{figure*}

In a new measurement of the \ttbar\ cross section in $e\mu$ final
states the top-quark mass is determined via the dependence of the
predicted cross section on the top-quark pole mass~\cite{pole_mass}. This
result extracts, for the first time, unambiguously the top-quark pole
mass since the experimental acceptance and therefore the extracted
cross section is basically constant
as a function of the assumed mass, and therefore the question of the
difference between extracted mass and pole mass becomes
insignificant. The result is presented in
Fig.~\ref{fig:properties}a. The extracted top-quark pole mass is
$m_t^{\rm pole} = 172.9^{+2.5}_{-2.6}$~GeV.  
At a future linear
collider such as the ILC, however, it is expected to measure a
well-defined top-quark mass at the \ttbar-production threshold with an uncertainty of
the order of 100~MeV~\cite{ilc} which will be very significant progress.

Many important and interesting measurements of other top-quark
properties in both \ttbar\ and single-top-quark production have been
presented at this conference~\cite{properties}. 
The results involve, for example,
measurements of \ttbar\ spin correlation (see
Fig.~\ref{fig:properties}b), top-quark polarization in \ttbar\
production (see Fig.~\ref{fig:properties}c) and in single-top-quark production (see
Fig.~\ref{fig:properties}d), $W$-boson helicity fractions extracted in top-quark 
decays in \ttbar\ and single-top-quark production (for the latter see
Fig.~\ref{fig:properties}e), searches for CP-violation, searches for
anomalous top-quark couplings and many more. All results agree with
the SM predictions. 

Deviations between measurement and prediction of the top-antitop-quark
forward-backward
asymmetry and the \ttbar\ charge asymmetry at the Tevatron have caused
lots of interest in our field in recent years. 
Similar analyses of the charge asymmetry in \ttbar\ events
at the LHC show agreement between data and prediction. Furthermore,
measurements of such asymmetries are performed in $b\bar{b}$ production at the
Tevatron and at the LHCb experiment and are in agreement with the SM
prediction, too. 
At the Tevatron the measurements were repeated with the full
dataset and the analyses were improved in accuracy with the
aim of increasing the potential to discover new physics. As a result
the new measurements in the lepton+jets and dilepton channels exploring the full
Tevatron dataset agree with recent SM
predictions~\cite{Bernreuther:2012sx}. This is summarized in
Fig.~\ref{fig:properties}f.

\section{Top quark interaction with the Higgs boson}
Because of the large mass of the top quark and since the largest
quantum corrections to the Higgs-boson mass involves a top-quark loop,
the top quark is believed to be closely related to the mechanism of
mass generation together with the Higgs boson. Thus, to test our
current theory of elementary particles in detail, it is of particular
importance to investigate top-quark interactions with the Higgs boson. 
At the LHC this can be studied in associated $\ttbar H$
production~\cite{tth}. Here, $H \to b\bar{b}$, $H \to \gamma\gamma$, $H \to
WW, ZZ$ (leading to multilepton final states) and $H \to \tau\tau$
final states are investigated with currently the most sensitive
channel being the first one. The information of many observables
sensitive to separate between signal and background is combined 
using multivariate analysis techniques, such as the construction of
neural networks (NN). An example NN output distribution in the
single-lepton $H \to b\bar{b}$ search from the ATLAS experiment is presented in
Fig.~\ref{fig:searches}a. This shows that good separation between
signal (concentrated at large values) and  background (concentrated at
low values) is achieved. 

A summary of all searches for $\ttbar H$ production from the CMS
experiment is shown in Fig.~\ref{fig:searches}b. Similar results exist from the
ATLAS Collaboration. All results are in agreement with the SM
prediction. The CMS Collaboration has found an excess above the
background-only expectation of 3.4 standard deviations. 
Furthermore, searches for $tHq$ production where the cross section can be
enhanced due to anomalous contributions to the top-Yukawa coupling are
performed~\cite{thq}. No hint for new physics is found.

\section{Searches for new physics in the top quark sector}
Due to the potentially special role of the top quark in electroweak
symmetry breaking, the top quark might be connected to physics beyond the
SM. New physics might show up in the data directly as bumps in mass
distributions but it could also be found through deviations from the
SM in high-precision top-quark-property measurements, for example, as
a consequence of anomalous couplings or due to the existence of new
Supersymmetric (SUSY) particles. At this conference, examples of both
direct searches for new physics studying kinematic properties and
indirect searches through high
precision measurements of top-quark properties were
presented.

\begin{figure*}[htpb!]
\vspace*{-3.7cm}
\begin{center}
\subfigure[]{\includegraphics[width=0.32\textwidth]{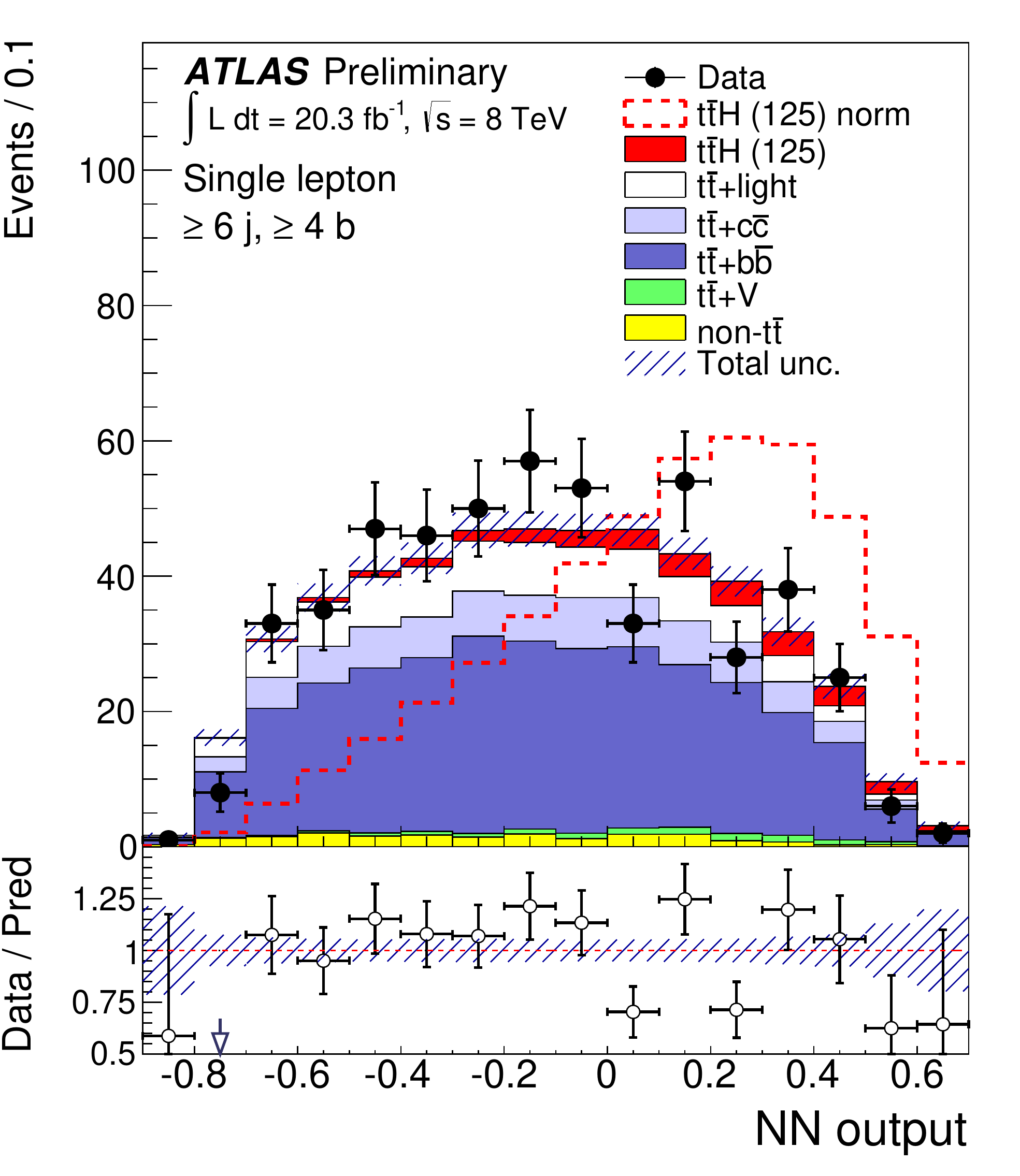}}\label{fig:Hbb_ATLAS}
\subfigure[]{\includegraphics[width=0.56\textwidth]{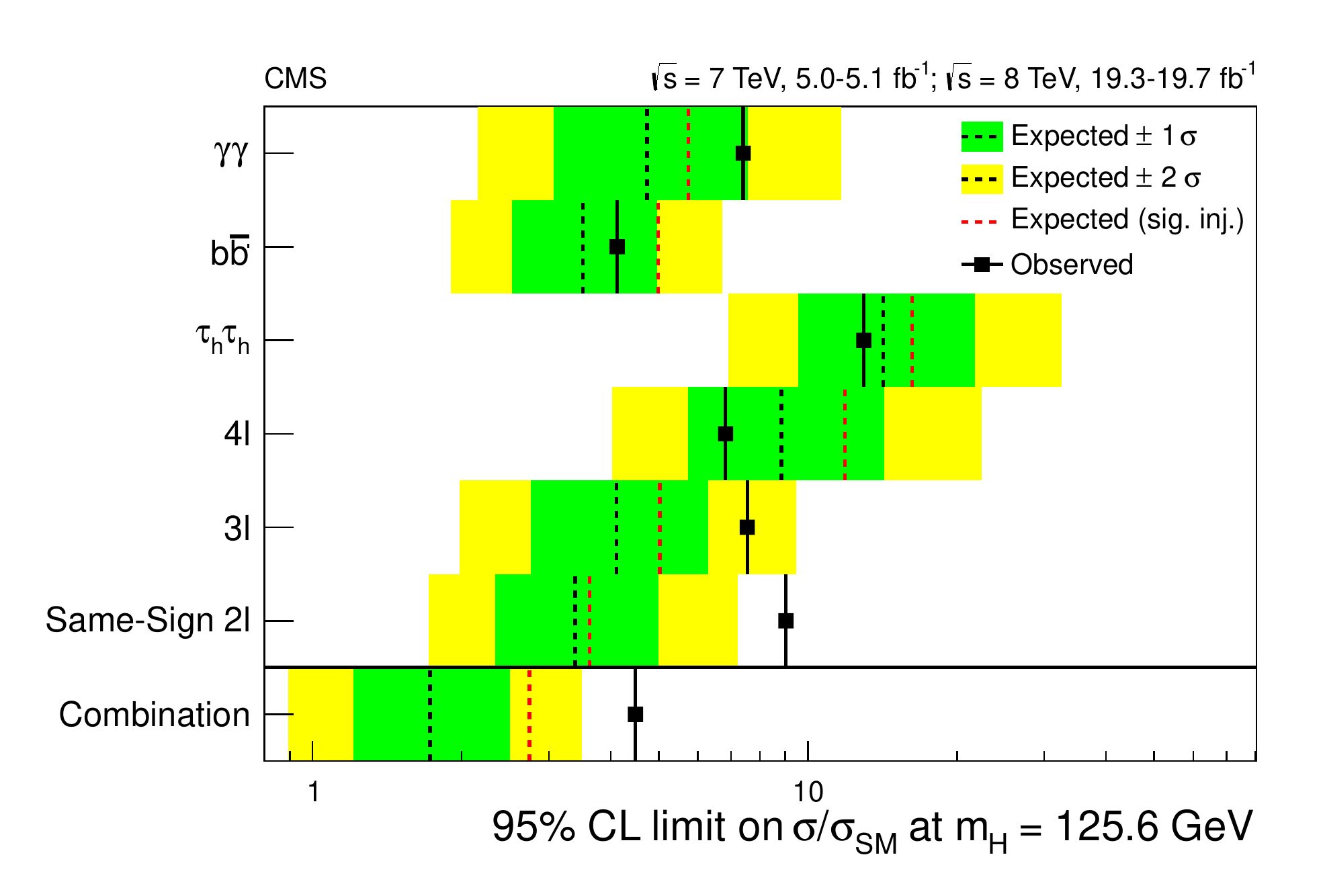}}\label{fig:ttH_sum_CMS}\\
\vspace*{-0.3cm}
\subfigure[]{\includegraphics[width=0.40\textwidth]{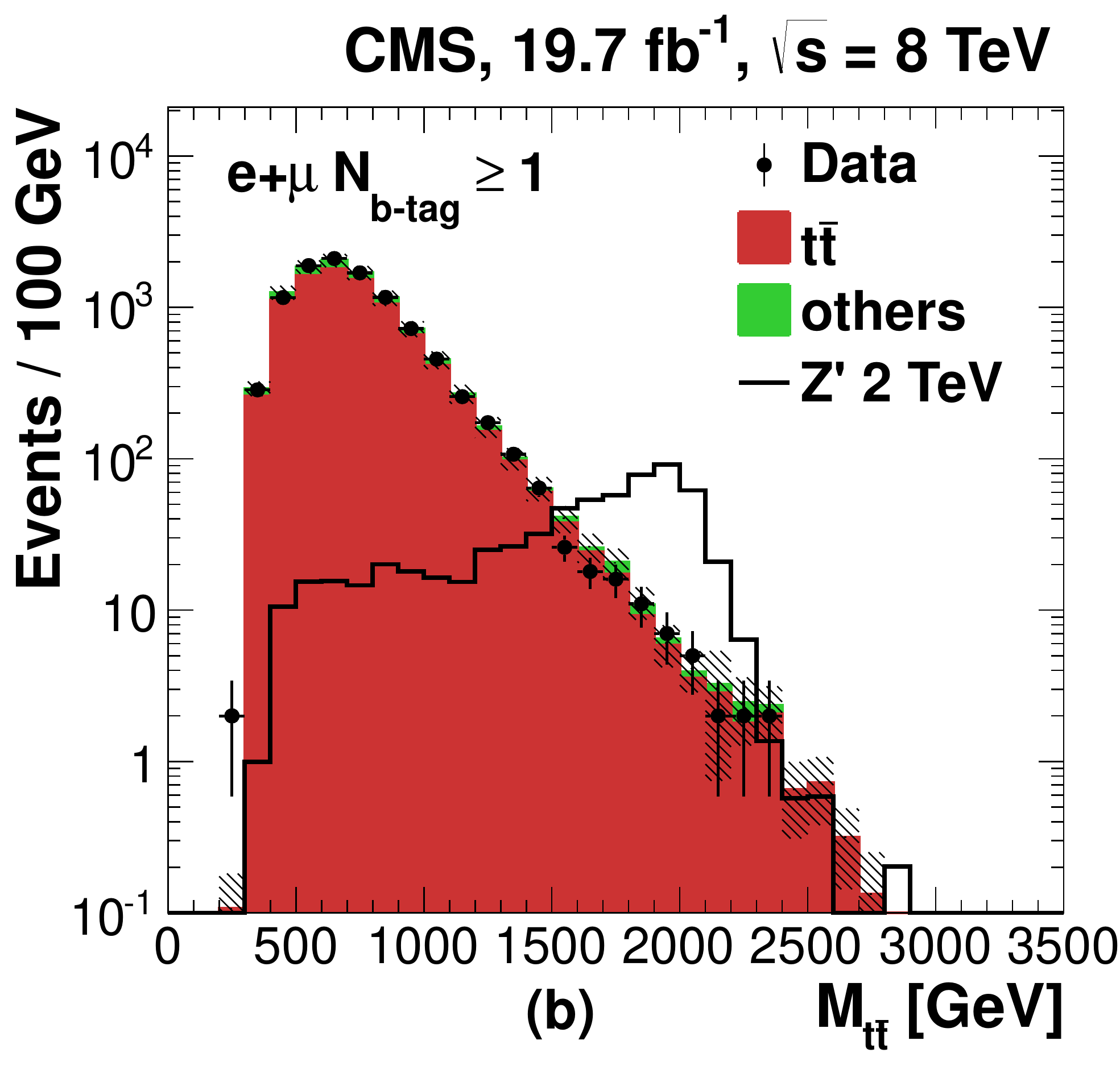}}\label{fig:ttres_CMS}\hspace{1.5cm}
\subfigure[]{\includegraphics[width=0.39\textwidth]{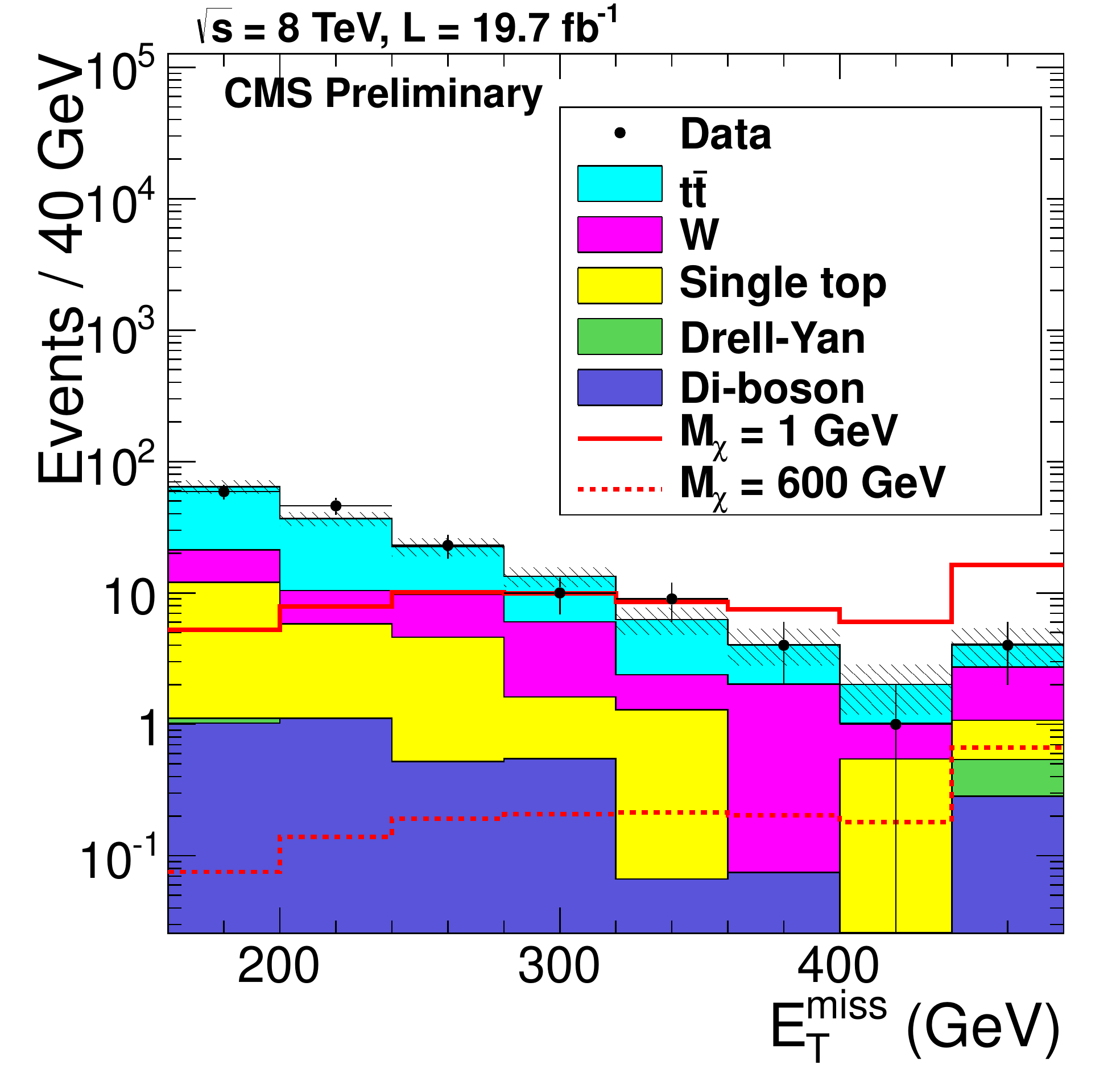}}\label{fig:DM_CMS}\\
\vspace*{-0.3cm}
\subfigure[]{\includegraphics[width=0.45\textwidth]{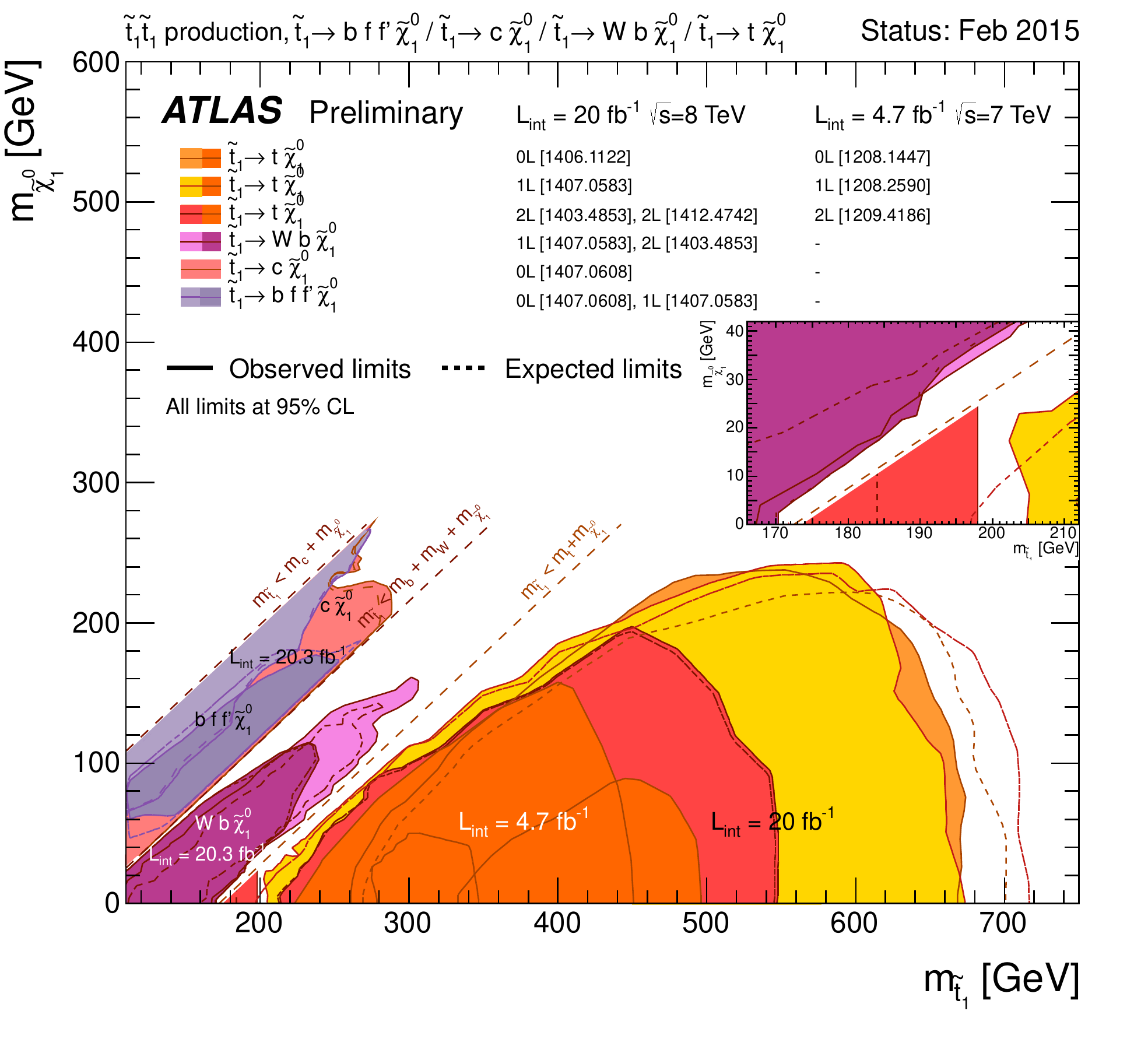}}\label{fig:stop_ATLAS}
\subfigure[]{\includegraphics[width=0.42\textwidth]{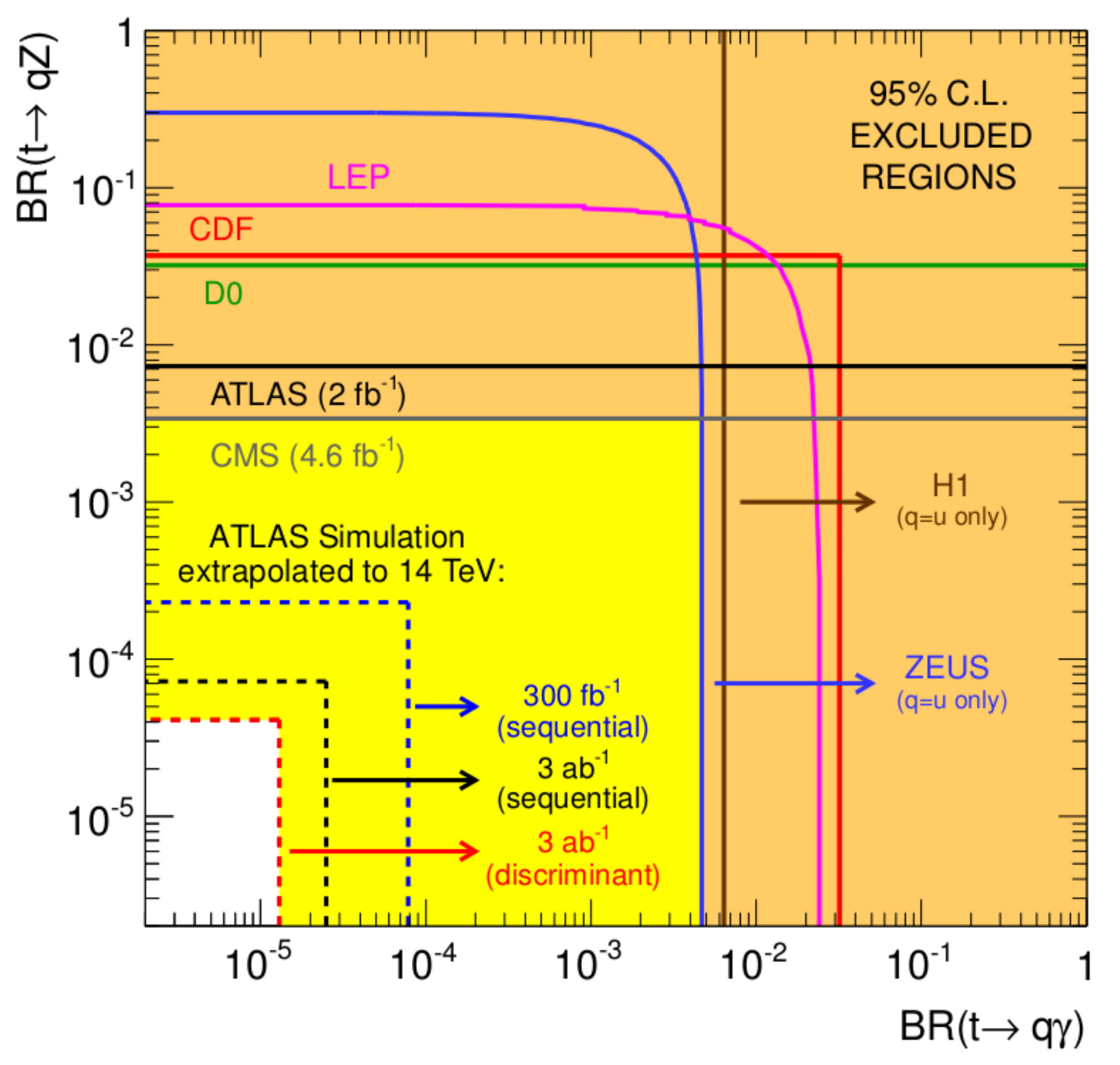}}\label{fig:FCNC}
\end{center}
\vspace*{-0.6cm}
\caption{Searches in the top-quark sector: (a) Example NN output
  distribution to search $\ttbar H$ production in single-lepton final states with at least six jets
  and at least four $b$-tagged jets at 8~TeV.  Data is compared to
  different backgrounds (colored histograms) and the 
  $\ttbar H$ signal (dashed line); (b) limit at the 95\% confidence
  level (CL) on the extracted $\ttbar H$ cross
section normalized by the SM prediction for different Higgs-boson
  decay channels combining data taken at 7 and 8 TeV;
(c) distribution of invariant \ttbar\ mass using reconstructed
using a scheme optimized for boosted top quarks to search for \ttbar\
resonances in lepton+jets final states. Data at 8 TeV are compared to 
backgrounds (colored histograms) and to a $Z' \to \ttbar$ production
signal example (solid line);
(d) missing transverse energy distribution in a search for dark matter
produced in association with a pair of top quarks. Data is compared to
background (colored histograms) and two examples of simulated
dark-matter signals with a mass of 1~GeV (solid line) and 600~GeV
(dotted line) and an interaction scale of 100 GeV;
(e) summary of exclusion limits for various modes of top-squark
decay. The red triangle illustrates the limit derived from the
spin-correlation measurement;
(f) observed limits at 95\% CL on the branching ratios in searches for FCNC $t \to qZ$ and $t
\to q\gamma$ decays at different collider experiments (solid
lines). Limits expected by ATLAS with 300~fb$^{-1}$ or 3~ab$^{-1}$
of integrated luminosity are added (dashed lines).}
\label{fig:searches}
\end{figure*}

For example, direct searches for \ttbar\ resonances and heavy $Z'$
bosons (see Fig.~\ref{fig:searches}c), heavy $W'$  bosons, vector-like   
heavy $T'$ and $B'$ quarks, excited top quarks, $tb$ resonances and dark matter (see
Fig.~\ref{fig:searches}d) were performed. No hint for new physics was
found~\cite{searches}. 

The LHC collaborations also perform a wide program of searching for
the SUSY scalar partner of the top quark, the top
squark~\cite{stop}. Large regions in the parameter space given by the top-squark mass
and the mass of the lightest SUSY particle (LSP) are excluded in direct searches exploiting
kinematic differences. If, however, the LSP (for example a
neutralino \xiz) is light and the top-squark mass is only slightly larger
than the top-quark mass, two-body decays $\stopi\rightarrow t\xiz$ in
which the momentum of \xiz\ is very small can
predominate. In this area the kinematics of top-squark-pair production event candidates
look very similar to those from
\ttbar\ production, and direct searches are therefore not
sensitive. Events with top-squark pairs, however, can be distinguished from
\ttbar\ events through an increase of the measured \ttbar\ cross
section, and since top squarks have zero spin, through measuring
angular correlations sensitive to spin correlation. As a result,
top-squark-pair production has not been detected, but a region for low
LSP masses and top-squark masses close to the top-quark mass are
excluded which is complementary to regions excluded by direct
searches~\cite{stop}. This is shown by the red triangle in
Fig.~\ref{fig:searches}e. This is an example of 
using a "SM candle" to probe for new physics beyond.

Finally, searches for anomalous couplings and flavor-changing neutral-current (FCNC)
couplings are of particular interest. No hint for the existence of
such couplings has been found so far~\cite{fcnc}. The respective LHC limits on FCNC branching ratios in $t
\to qZ$ and $t \to q\gamma$ decays are 
complementary to results from the LEP $e^+e^-$ collider, the HERA $ep$
collider and the Tevatron \ppbar\ collider as can be seen in Fig.~\ref{fig:searches}f.
Typical models beyond the SM may give rise to FCNC branching ratios at
the level of ${\rm BR}<10^{-4}$. To be sensitive enough to probe such
models a High Luminosity LHC~\cite{hl-lhc} or a future circular
$e^+e^-$ collider~\cite{fcc-ee}, for example, would be needed.

\section{Conclusions}
Many exciting results covering a very broad spectrum of particle physics
research were presented at the
TOP 2014 Conference. Particular highlights were high
precision measurements of top-quark production cross sections, the top-quark
mass, top-quark couplings and many more top-quark
properties. Furthermore, the developments of high precision theory
tools, such as NLO+multileg MC simulations and even first calculations
at NNLO QCD are just breathtaking, and clearly the experimental
analyses profit from this significantly~\cite{theory}. There are vast improvements in
understanding theoretical uncertainties, combining results and
developing tools such as top-tagging algorithms. This allows us, for
example, to perform high-precision tests of QCD using unfolded
differential distributions 
measured in fiducial regions of the detector that can directly be
compared to theoretical calculations. 
Many new 
top-quark property measurements exist now utilizing both single-top-quark
and \ttbar\ production events. To understand the current 
differences in top-quark-mass measurements of different experiments is 
an important challenge to perform in the near future. 
Moreover, new SM processes were
observed for the first time such as
$s$-channel single-top-quark production and $\Wt$ production. Evidence
was found for SM $\ttbar Z$, $\ttbar W$ and 
$\ttbar H$ production. Furthermore,
sensitive searches for new physics in the top-quark sector are
performed in complementary ways, by traditional direct searches
exploring kinematic properties and also by performing high-precision 
measurements of top-quark properties. However, so far all results agree with
the SM predictions.

At the LHC Run-II, top-quark physics will become even more exciting,
with more SM processes to be observed, many high-precision property
and cross-section measurements to be performed, electroweak and
Yukawa couplings to be extracted, and hopefully new
physics to be discovered and explored.

\section*{Acknowledgments}
I thank the organizers of Top 2014 for this exciting,
very well-organized conference at this beautiful place at the C\^{o}te
d'Azur, the speakers for their excellent
presentations and the conveners for their ability to create an
inspiring discussion atmosphere. I am grateful that I was honored to give
the experimental summary presentation of the conference representing
so many brilliant results. I am thankful to The Royal Society (UK) for the support of
my research.



\section*{References}
\begin{thebibliography}{99}



\bibitem{ttbar_xsec} E.~Shabalina, these proceedings (2014);
  J.~Brochero-Cifuentes, these proceedings (2014).

\bibitem{xsec_tt} 
  M.~Czakon, P.~Fiedler and A.~Mitov,
  Phys.\ Rev.\ Lett.\  {\bf 110}, 252004 (2013).

\bibitem{xsec_diff} C. Diez-Pardos, these proceedings (2014); J.Katzy, these proceedings (2014).

\bibitem{sitop_tev} M.~Ronzani, these proceedings (2014).

\bibitem{sitop_lhc} A.~Jafari, these proceedings (2014).

\bibitem{sitop-kidonakis}
%
  N.~Kidonakis,
  Phys.\ Rev.\ D {\bf 83}, 091503 (2011);
 %
  N.~Kidonakis,
  Phys.\ Rev.\ D {\bf 82}, 054018 (2010);
%
  N.~Kidonakis,
  Phys.\ Rev.\ D {\bf 81}, 054028 (2010).

\bibitem{ttZ_ttW} T.~ Vazquez-Schroeder, these proceedings (2014).

\bibitem{systs} M.~J.~Costa, these proceedings (2014); M. Seidel, these proceedings (2014).

\bibitem{combi_lhc} G.~Cortiana, these proceedings (2014).

\bibitem{combi_world} Y. Peters, these proceedings (2014).

\bibitem{tools} E.~Usai, these proceedings (2014).

\bibitem{boosted} J. Erdmann,
  these proceedings (2014).

\bibitem{mass_d0} V.~Sharyy, these proceedings (2014).

\bibitem{mass_cms} E.~Schlieckau, these proceedings (2014).

\bibitem{pole_mass} S. Adomeit, these proceedings (2014).

\bibitem{ilc} F. Simon, these proceedings (2014).

\bibitem{properties} A. Jung, these proceedings (2014); R.~Hawkings,
  these proceedings (2014).

\bibitem{Bernreuther:2012sx} 
  W.~Bernreuther and Z.~G.~Si,
  Phys.\ Rev.\ D {\bf 86}, 034026 (2012).

\bibitem{tth} M.~Owen, these proceedings (2014).

\bibitem{thq} C.~B{\"o}ser, these proceedings (2014).

\bibitem{searches} F.~Canelli, these proceedings (2014).

\bibitem{stop} T.~Eifert, these proceedings (2014).

\bibitem{fcnc} R. Goldouzian, these proceedings (2014).

\bibitem{hl-lhc} P.~Ferreira da Silva, these proceedings (2014).

\bibitem{fcc-ee} B.~Fuks, these proceedings (2014).

\bibitem{theory} T.~M.~P.~Tait, these proceedings (2014), and
  references therein.


\end{thebibliography}
\end{document}